\documentclass[preprint,12pt]{elsarticle}

\usepackage{moreverb,url}
\usepackage{bm}
\usepackage[colorlinks,bookmarksopen,bookmarksnumbered,citecolor=red,urlcolor=red]{hyperref}
\usepackage{amsmath,amssymb}
\usepackage{graphicx,caption,setspace}
\newcommand{\mse}{\xi} 

\journal{}

\begin{document}

\begin{frontmatter}

%% Title, authors and addresses

%% use the tnoteref command within \title for footnotes;
%% use the tnotetext command for theassociated footnote;
%% use the fnref command within \author or \address for footnotes;
%% use the fntext command for theassociated footnote;
%% use the corref command within \author for corresponding author footnotes;
%% use the cortext command for theassociated footnote;
%% use the ead command for the email address,
%% and the form \ead[url] for the home page:
%% \title{Title\tnoteref{label1}}
%% \tnotetext[label1]{}
\author{David Benkeser}
\ead{benkeser@emory.edu}
\author{Andrew Mertens}
\author{Benjamin F. Arnold}
\author{John M. Colford, Jr.}
\author{Alan Hubbard}
\author{Aryeh Stein}
\author{N. L’ntshotshol{\`e} Jumbe}
\author{Mark J. van der Laan}
%% \ead[url]{home page}
%% \fntext[label2]{}
%% \cortext[cor1]{}
%% \address{Address\fnref{label3}}
%% \fntext[label3]{}

\title{A machine learning-based approach for estimating and testing associations with multivariate outcomes}

%% use optional labels to link authors explicitly to addresses:
%% \author[label1,label2]{}
%% \address[label1]{}
%% \address[label2]{}

\begin{abstract}
We propose a method for summarizing the strength of association between a set of variables and a multivariate outcome. Classical summary measures are appropriate when linear relationships exist between covariates and outcomes, while our approach provides an alternative that is useful in situations where complex relationships may be present. We utilize ensemble machine learning to detect nonlinear relationships and covariate interactions and propose a measure of association that captures these relationships. A hypothesis test about the proposed associative measure can be used to test the strong null hypothesis of no association between a set of variables and a multivariate outcome. Simulations demonstrate that this hypothesis test has greater power than existing methods against alternatives where covariates have nonlinear relationships with outcomes. We additionally propose measures of variable importance for groups of variables, which summarize each groups' association with the outcome. We demonstrate our methodology using data from a birth cohort study on childhood health and nutrition in the Philippines. 
\end{abstract}


\end{frontmatter}

\section{Introduction}

Neurocognitive impairment may affect 250 million children under five years globally.\cite{black2017} The Healthy Birth, Growth, and Development knowledge integration initiative was established, in part, to inform global public health programs toward optimizing neurocognitive development.\cite{jumbe2016} Neurocognitive development is often studied through studies that enroll pregnant women and follow their children through early childhood and adolescence. Covariate information about the child's parents, environment, and somatic growth is recorded at regular intervals and in early adolescence, children complete tests that measure diverse domains of neurocognitive development such as motor, mathematics, and language skills. Researchers are often interested in assessing the correlation between covariate information and neurocognitive development. Such an assessment may be useful for developing effective prediction algorithms that identify children at high risk for neurocognitive deficits.\cite{walker2007child} 

A common approach for describing the strength of association between covariates and a multivariate outcome is canonical correlation analysis.\cite{hotelling1935most,hotelling1936relations} Canonical correlation maximizes the correlation between a linear combination of the multivariate outcome and a linear combination of the covariates. Several test statistics have been proposed for significance testing of canonical correlation including Wilks' $\Lambda$ \cite{wilks1932certain}, the Hotelling-Lawley trace \cite{hotelling1935most,hotelling1936relations}, the Pillai-Bartlett trace \cite{bartlett1941statistical,pillai1956distribution}, and Roy's largest root \cite{roy1945individual}. One potential drawback to canonical correlation analysis is that it may fail to identify associations when nonlinear relationships exist between covariates and outcomes. This has led to recent interest in nonlinear extension of canonical correlation analysis \cite{nandy2003novel,andrew2013deep,michaeli2016nonparametric}.

Another common approach in settings with multivariate outcomes is latent variable analysis. Many definitions of latent variables appear in the literature, and we refer interested readers to more thorough discussions in \cite{glymour1987discovering,sobel1994,edwards2000nature,hagglund2001,bollen2002}. A commonly used definition is that a latent variable is a hypothetical construct that cannot be measured by the researcher and that describes some underlying characteristic of the participant. Others have strongly rejected the use of latent variables, instead opting for empirical explanations of observed phenomenon \cite{skinner1976behaviorism}. For our purposes it suffices to say that latent variable analysis tends entail an unsupervised grouping of the observed outcomes into a set of lower-dimensional latent features. One technique commonly employed to this end is principal components analysis (PCA). Researchers often use PCA to reduce the observed multivariate outcome to a low-dimensional (e.g., univariate) outcome and may examine the factor loadings to ascribe a-posteriori meaning to the reduced outcomes. Researchers further may use these outcomes to test for associations of covariates. However, such tests may fail to identify associations between predictors and outcomes due to the unsupervised nature of the outcomes' construction.

In this work, we propose an alternative method for measuring association between a multivariate outcome and a set of covariates. Our approach is related to canonical correlation analysis, but rather than maximizing the correlation between linear functions of covariates and outcomes, we maximize the predictive accuracy of a flexible machine learning algorithm and a convex combination of the outcome. The method identifies the univariate outcome that is ``most easily predicted'' by the covariates and a summary measure of how well that ``easiest-to-predict'' outcome may be predicted. This approach adapts to complex relationships between covariates and outcomes and identifies associations in situations where traditional canonical correlation does not. However, in contrast to recent nonparametric canonical correlation proposals, our method provides a measure of association on a familiar scale, and also provides asymptotically justified inference, including confidence intervals and hypothesis tests. 

In certain situations, our method may further provide a novel means of constructing a clinically interpretable latent variable. For example, in studies of neurocognitive development with measured genetic information, our method could be used to identify a univariate measure of neurocognitive development that reflects the components of neurocognitive development most strongly associated with genetics. This univariate outcome may be used to represent the latent trait of ``heritable intelligence,'' and could be used to study associations and gene/environment interactions. Nevertheless, we view our method primarily as a latent variable approach only in the sense of Harman (1960) \cite{harman1960modern}, who described latent variables as a convenient means of summarizing a number of variables in many fewer factors. Our approach fits in this definition by providing a single summary measure of the strength of association between covariates and multivariate outcomes. 

\section{Defining the target parameter}

Suppose we observe $n$ independent copies of $O := (X_1,\dots,X_D,Y_1,\dots,Y_J)$, where $X := (X_1,\dots,X_D)$ is a $D$-dimensional vector of covariates and $Y := (Y_1,\dots,Y_J)$ is a $J$-dimensional vector of continuously-valued outcomes. Without loss of generality, we assume that each outcome has mean zero and standard deviation one and otherwise make no assumptions about the true distribution $P_0$ of $O$. We define $Y_{\omega} := \sum_{j=1}^J \omega_j Y_j$ as a convex combination of the outcomes, where $\omega_j \ge 0$ for all $j$ and $\sum_j \omega_j = 1$. We are interested in finding $\omega_0$, the weights that make $Y_{\omega_0}$ ``easiest-to-predict'' based on $X$. We formalize this notion presently. Subsequently, a summary measure of how well $Y_{\omega_0}$ can be predicted using $X$ will serve as a summary measure of the strength of the association between $X$ and $Y$. Rejecting a null hypothesis of zero association allows us to conclude that there is an association between $X$ and at least one of the components of $Y$. 

To formalize, suppose we are given weights $\omega$ and a function $\psi_{0,\omega}$ that takes $x$ as input and outputs $\psi_{0,\omega}(x)$, a real-valued prediction of $Y_\omega$. In the sequel, we discuss how one might construct such a function using the observed data. A measure of how well $\psi_{0,\omega}$ predicts $Y_\omega$ is mean squared-error, $\mse_{0,\omega}(\psi_{0,\omega}) := E_0[\{Y_\omega - \psi_{0,\omega}(X)\}^2]$, which measures the average squared distance between the prediction made by $\psi_{0,\omega}$ and the outcome $Y_{\omega}$. Here we use the notation $E_0\{f(X,Y)\}$ to denote the average value of $f(X,Y)$ under the true distribution $P_0$. Mean squared-error can be scaled to obtain a measure of accuracy that does not depend on the scale of the outcome. Specifically, let $\mu_{0,\omega} := E_0(Y_\omega)$ denote the marginal mean of $Y_j$ and define \begin{equation}
\rho^2_{0,\omega}(\psi_{0,\omega}) := 1 - \frac{\mse_{0,\omega}(\psi_{0,\omega})}{\mse_{0,\omega}(\mu_{0,\omega})} \ , \label{trueR2}
\end{equation}
as the proportional reduction in mean squared-error when using $\psi_{0,\omega}$ as opposed to $\mu_{0,\omega}$ to predict $Y_\omega$. We refer to this as a nonparametric $R^2$, as this measure may be interpreted similarly as the $R^2$ measure in the familiar context of linear models: values near to one indicate that nearly all of the variance in $Y_\omega$ is explained by $\psi_{0,\omega}$. However, we note that the squared notation here is a misnomer in that $\rho^2_{0,\omega}(\psi_{0,\omega})$ can be negative, which indicates that the marginal mean of $Y_\omega$ provides a better fit than $\psi_{0,\omega}$. Nevertheless, we maintain the colloquial squared-notation.

We define $\omega_0$ as the set of weights that is optimal with respect to nonparametric $R^2$, $\omega_0 := \mbox{argmax}_\omega \rho^2_{0,\omega}(\psi_{0,\omega})$. We assume that such a maximal value exists and is unique. The value $\rho^2_{0,\omega_0}(\psi_{0,\omega_0})$ describes how well the optimally combined outcome may be predicted. This measure provides a summary of the association between $X$ and $Y$ that is closely related to canonical correlation \cite{hotelling1936relations}. Indeed, if $\psi_\omega$ is based on a linear combination of $X$, then $\rho^2_{0,\omega_0}(\psi_{\omega_0})$ is nearly identical to the squared correlation of the first canonical variates. However, canonical correlation reflects only the strength of linear associations between $X$ and $Y$, while $\psi_{0,\omega_0}$ may include nonlinear functions of $X$. Thus, $\rho^2_{0,\omega_0}(\psi_{0,\omega_0})$ may more accurately summarize the association between $X$ and $Y$ in situations where relationships between covariates and outcomes are not well described by linear models. Our goals are thus: (i) find a suitable prediction function for a convex combination of the multivariate outcome; (ii) approximate the optimal weights $\omega_0$; (iii) summarize the strength of the association between the prediction function and the optimally weighted outcome. 

Restricting $\psi_{0,\omega}$ to linear functions of $X$, as in canonical correlation analysis, enables simultaneous optimization over choice of prediction function and outcome weights. Furthermore, if $X$ and $Y$ are of relatively low dimension, then canonical correlation may provide a reasonable measure of association between $X$ and $Y$. However, when considering more complex functions of $X$, simultaneous optimization appears to be a much more difficult task. Furthermore, when more aggressive data fitting approaches are employed, overfitting becomes concern and may prevent simple correlative statistics from being used to summarize associations. To overcome these difficulties, we propose to use an approach centered around cross-validation. For goal (i), we develop a prediction function for each component of $Y$ using a cross-validation-based ensemble machine learning technique, which ensures that the relationship between $X$ and $Y$ is captured as accurately as possible.  For goal (ii), we propose to maximize a cross-validated performance measure in order to determine outcome weights. For goal (iii), we propose to cross-validate the entire procedure to obtain a summary of the association between $X$ and $Y$. We propose closed-form variance estimators for this associative measure that can be used to generate confidence intervals and hypothesis tests.

\section{Methodology}
% If instead, we have a prediction function $\psi_0$ available, for example, based on external data, then we may define an empirical version of (\ref{trueR2}) \begin{equation} 
% \rho^2_{n,\omega}(\psi_\omega) := 1 - \frac{\sum_{i=1}^n \{Y_{\omega,i} - \psi_{0,\omega}(X_i)\}^2}{\sum_{i=1}^n (Y_{\omega,i} - \mu_{n,\omega})^2} \ , \label{empiricalR2}
% \end{equation}
% where $\mu_{n,\omega} := \frac{1}{n}\sum_{i=1}^n Y_{\omega,i}$ and obtain an estimate of $\omega_0$ as $\mbox{argmax}_\omega \rho^2_{n,\omega}(\psi_{0,\omega})$. We may use $\rho^2_{n,\omega_n}(\psi_{0,\omega_n})$ as summary measure of the association between $X$ and $Y$. However, in the more realistic scenario in which an external estimate of $\psi_{0,\omega}$ is not available, obtaining a reasonable estimate of $\rho^2_{n,\omega_n}(\psi_{0,\omega_n})$ is more difficult. 

\subsection{Super learning}
First, we note that given $\omega$, the maximizer of $\rho^2_{0,\omega}(\psi_\omega)$ over all functions mapping an observation $x$ to a real value is $\psi_{0,\omega} := E_0(Y_\omega \ | \ X)$. Due to linearity of expectation, \begin{equation}
E_0(Y_\omega \ | \ X) = E_0\biggl(\sum_{j=1}^J \omega_j Y_j \biggr) = \sum_{j=1}^J \omega_j E_0(Y_j \ | \ X) \ .  \label{linearExpectation}
\end{equation}
This implies that for any $\omega$, a prediction function for $Y_\omega$ may be constructed by building a prediction function for each $Y_j$ and weighting those prediction function by $\omega$. An alternative approach is to use multivariate regression methods simultaneously predict the whole vector $Y$.\cite{izenman2008modern} However, we find the univariate approach appealing in that it allows a rich collection of univariate learning methods to be utilized. In particular, we propose to use regression stacking \cite{wolpert1992stacked,breiman1996stacked}, also known as super learning \cite{vdlpolley:2007:statappgenetics}, to construct a prediction function for each outcome. Super learning entails constructing a library of $M_j$ candidate learners for each of the $J$ outcomes. The library of estimators can include parametric model-based estimators as well as machine learning algorithms. Parametric model-based methods include linear regression of $Y_j$ on $X$, while machine learning-based methods include random forests \cite{breiman2001random}, gradient boosted machines \cite{friedman2001greedy}, and deep neural networks \cite{ripley2007pattern}. If $X$ is high-dimensional, learners can incorporate dimension-reduction strategies. For example, we may include a main-terms linear regression of $Y_j$ on all variables in $X$, while also including a linear regression using only a subset of $X$ chosen based on their univariate association with $Y_j$. The library of learners can also include different choices of tuning parameters for machine learning algorithms, an important consideration for methods that require proper selection of many tuning parameters (e.g., deep learning). 

Because there is no way to know a-priori which of these myriad estimators will be best, a cross-validated empirical criterion is used to adaptively combine, or ensemble, the various learners. This approach is appealing in the present application in that the best learner for each outcome might be different. By allowing the data to determine the best ensemble of learners for each outcome, we expect to obtain a better fit over all outcomes and thereby obtain a more accurate summary of the association between $X$ and $Y$. Indeed, the super learner is known to be optimal in the sense that, under mild assumptions, its goodness-of-fit is essentially equivalent to the (unknown) best-fitting candidate estimator.\cite{van2004asymptotic,vanderLaan&Dudoit&vanderVaart06} A full treatment of super learning\cite{vdlpolley:2007:statappgenetics} and R package\cite{superlearnerpackage} are available. Here, we provide a brief overview of the procedure. 

\begin{figure}
\centering
\includegraphics[width=0.7\textwidth]{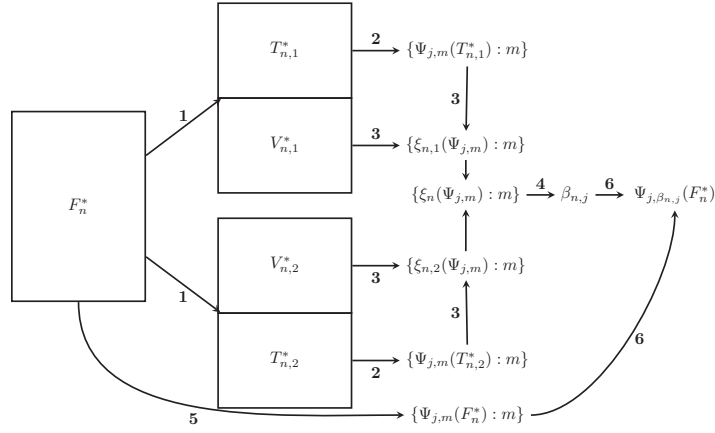}
\caption{A flowchart illustrating a two-fold super learner procedure, which maps a set of observed data indices, $F_n^*$ into $\Psi_{j,\beta_{n,j}}(F_n^*)$ an ensemble prediction function for outcome $Y_j$. Notation: $T_{n,k}$, the $k$-th training sample; $V_{n,k}$, the $k$-th validation sample; $\Psi_{j,m}(T_{n,k})$, the $m$-th learner for the $j$-th outcome fit in the $k$-th training sample; $\mse_{n,k}(\Psi_{j,m})$, the $k$-th cross-validated mean squared-error; $\mse_{n}(\Psi_{j,m})$, cross-validated mean squared-error; $\beta_{n,j}$, a $J$-length vector of super learner weights; $\Psi_{j,m}(F_n^*)$, the $m$-th learner for the $j$-th outcome fit using $\{O_i : i \in F_n^*\}$.}
\label{sl_flowchart}
\end{figure}

Super learning is often based on $K^*$-fold cross validation, and we illustrate the method for $K^* = 2$ in Figure \ref{sl_flowchart}. The super learner for a given outcome $Y_j$ may be implemented in the following steps:

\noindent \textbf{Step 1: Partition the data. }
Starting with a data set consisting of observations $\{O_i : i \in F_n^*\}$ for some $F_n^* \subseteq \{1,\dots,n\}$, we randomly partition the data into $K^*$ splits of approximately equal size. We define the $k$-th training sample as the observed data with observations in the $k$-th split removed. We use $T_{n,k}^* \subset F_n^*$ to denote the indices of observations in the $k$-th training sample and $V_{n,k}^*$ to denote the indices of observations in the $k$-th validation sample for $k=1,\dots,K$. 

% and subsequently using observations in the validation sample to evaluate performance.

\noindent \textbf{Step 2: Fit learners in training data. }
We now use the observations in the training sample to fit each of the $M_j$ candidate learners. We use $\Psi_{j,m}$ to denote the $m$-th candidate learner for the $j$-th outcome. The learner $\Psi_{j,m}$ takes as input a set of indices and returns a prediction function for $Y_j$. We use $\Psi_{j,m}(T_{n,k})$ to denote the $m$-th candidate learner for the $j$-th outcome fit in the $k$-th training sample. For example, $\Psi_{j,m}$ might correspond with fitting an ordinary least squares linear regression of $Y_j$ on $X$, in which case $\Psi_{j,m}(T_{n,k})$ corresponds to the fitted linear predictor based on the regression fit in the $k$-th training sample and $\Psi_{j,m}(T_{n,k})(x)$ corresponds to the linear predictor of the fitted regression evaluated at the observation $x$.

\noindent \textbf{Step 3: Evaluate learners in validation data. }
Next, we use the data in the $k$-th validation sample to evaluate each of the learners $\Psi_{j,m}(T_{n,k}), \ m = 1,\dots,M_j$ via $$
\mse_{n,k}(\Psi_{j,m}) := \frac{1}{|V_{n,k}|} \sum_{i \in V_{n,k}} \{Y_{j,i} - \Psi_{j,m}(T_{n,k})(X_i) \}^2 \ , 
$$
where we use $|V_{n,k}|$ to denote the number of observations in the $k$-th validation sample. We obtain an overall estimate of the performance of the $m$-th learner via $$
\mse_{n}(\Psi_{j,m}) := \frac{1}{K} \sum\limits_{k=1}^{K} \mse_{n,k}(\Psi_{j,m}) \ . 
$$

\noindent \textbf{Step 4: Find ensemble weights. }
The cross-validation selector $\tilde{m}$ is defined as the single learner with the smallest $\mse_{n}$ and one may use $\Psi_{j,\tilde{m}}(F_n^*)$ as the learner for the $j$-th outcome, i.e., we refit the cross-validation-selected learner using all the data and use this as prediction function. However, it is often possible to improve on the fit of the cross-validation selector by considering ensemble learners of the form $\Psi_{j,\beta} = \sum_m \beta_{j,m} \Psi_{j,m}$, where the weights $\beta_{j,m}$ are non-negative for all $m$ and sum to 1. The prediction function $\Psi_{j,\beta_j}$ takes as input $X$, computes the prediction of $Y_j$ using each of the $M_j$ learners, and returns a convex combination of these predictions. We can compute $\beta_{n,j} := \mbox{argmin}_{\beta_j} \mse_n(\Psi_{j,\beta_j})$, the choice of weights that minimizes cross-validated mean squared-error, using standard optimization software. 

\noindent \textbf{Step 5: Refit learners using full data. }
We next refit the learners that received non-zero weight in $\beta_{n,j}$ using all observations in $F_n^*$. We denote these fits $\Psi_{j,m}(F_n^*)$ for $m=1,\dots,M_j$. 

\noindent \textbf{Step 6: Combine learners into super learner. }
The super learner is the convex combination of the $M_j$ learners using weights $\beta_{n,j}$ and is denoted by $\Psi_j(F_n^*) := \Psi_{j,\beta_{n,j}}(F_n)$. 

% Given $\omega$, the $J$ super learners may be combined as $\Psi_{\omega}(F_n) := \sum_j \omega_j \Psi_j(F_n)$ to obtain a learner for $Y_\omega$.

We conclude this subsection by noting that, while the super learner is an appealing choice in many applications, the remainder of our proposal does not rely on the super learner being used to develop a learner for $Y_\omega$. In the remainder, we often refer to $\{\Psi_j : j=1,\dots,J\}$ as being a set of super learners; however, these could just as well be any learning algorithm.

\subsection{Choosing weights for composite outcome}
\begin{figure}
\centering
\includegraphics[width = 0.7\textwidth]{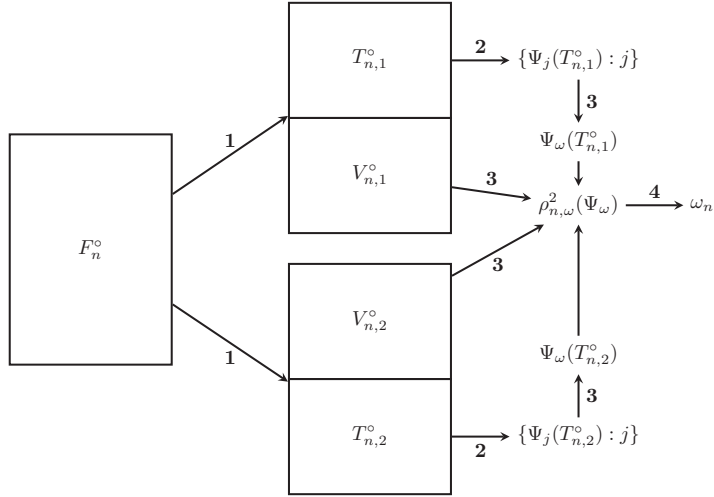}
\caption{A flowchart illustrating two-fold outcome weighting procedure, which maps a set of observed data indices, $F_n^\circ$ into $\Omega_n(F_n^\circ)$, a vector of convex weights for the outcome $Y$. Notation: $T_{n,k}^\circ$, the $k$-th training sample; $V_{n,k}^\circ$, the $k$-th validation sample; $\Psi_j(T_{n,k}^\circ)$, the super learner for the $j$-th outcome fit in the $k$-th training sample; $\Psi_\omega(T_{n,k})$, the composite super learner based on weights $\omega$; $\rho^2_{n,\omega}(\Psi_{\omega})$, the cross-validated nonparametric $R^2$ for the composite super learner.}
\label{weights_figure}
\end{figure}

The super learner is an aggressive learning algorithm and as such we must be concerned about overfitting when selecting weights for the composite outcome. If the super learner overfits the $j$-th outcome, then $Y_j$ may be erroneously upweighted in $Y_\omega$. To avoid overfitting, we propose to use an additional layer of $K^{\circ}$-fold cross-validation when selecting outcome weights. Figure \ref{weights_figure} illustrates our proposal for $K^{\circ} = 2$. 

\noindent \textbf{Step 1: Partition the data. }
Starting with a data set consisting of observations $\{O_i : i \in F_n^\circ\}$ for some $F_n^\circ \subseteq \{1,\dots,n\}$, we randomly partition the data into $K^\circ$ splits of approximately equal size. As above, we use $T_{n,k}^\circ \subset F_n^\circ$ to denote the indices of observations in the $k$-th training sample and $V_{n,k}^\circ$ to denote the indices of observations in the $k$-th validation sample for $k=1,\dots,K$.

\noindent \textbf{Step 2: Fit super learner for each outcome in training data. }
For each split $k$, we use the training sample to fit each outcome, resulting in learner fits $\Psi_j(T^{\circ}_{n,k})$ for $j=1,\dots,J$. We note that the super learner itself requires cross-validation so that this step will include nested cross-validation. That is, we repeat the super learning procedure outlined in the previous section with $F_n^* = T_{n,k}^\circ$ for $k=1,\dots,K^\circ$. 

\noindent \textbf{Step 3: Compute cross-validated $R^2$. }
%After training the learner in each of the $K^{\circ}$ splits, we obtain our estimate of $\omega_0$ as follows. 
For a given $\omega$, we can compute the super learner prediction function for $Y_\omega$, $\Psi_{\omega}(T^\circ_{n,k}) := \sum_{j=1}^J \omega_j \Psi_j(T^\circ_{n,k})$. For any $\omega$, we can use data in the $k$-th validation sample to compute an estimate of mean squared-error for $\Psi_{\omega}(T^\circ_{n,k})$,
 $$
\mse_{n,\omega,k}(\Psi_{\omega}(T^{\circ}_{n,k})) := \frac{1}{|V^{\circ}_{n,k}|} \sum_{i \in V^{\circ}_{n,k}} \{ Y_{\omega, i} -  \Psi_{\omega}(T^{\circ}_{n,k})(X_i) \}^2 \ . 
$$ 
The cross-validated mean squared-error is the sum over the validation folds, $\mse_{n,\omega}(\Psi_{\omega}) := \frac{1}{K} \sum_{k=1}^{K^{\circ}} \mse_{n,\omega,k}(\Psi_{\omega}). 
$ 
Similarly, we define $$ 
\bar{\Psi}_{\omega}(F_n^\circ) := \frac{1}{|F_n^\circ|} \sum_{i \in F_n^\circ} Y_{\omega,i} \ \ \ \mbox{and} \ \ \ \mse_{n,\omega}(\bar{\Psi}_{\omega}) := \frac{1}{|F_n^\circ|}\sum_{i \in F_n^\circ} \{Y_{\omega,i} - \bar{\Psi}_{\omega}(F_n^\circ)\}^2 \ , 
$$
as the empirical average weighted outcome among observations in $F_n^\circ$ and the empirical mean squared-error for predicting the composite outcome $Y_{\omega}$ using the composite sample means $\bar{\Psi}_{\omega}$. For any $\omega$, we can combine these elements to compute the cross-validated nonparametric $R^2$ measure as $$
\rho^2_{n,\omega}(\Psi_{\omega}) := 1 - \frac{
	\mse_{n,\omega}(\Psi_{\omega})
}{
	\mse_{n,\omega}(\bar{\Psi}_{\omega}) 
} \ .
$$

\noindent\textbf{Step 4: Maximize cross-validated $R^2$. }
The maximizer $\omega_n := \mbox{argmax}_{\omega} \rho^2_{n,\omega}(\Psi_{\omega})$ of nonparametric $R^2$ may be computed using standard optimization software. In our simulation and data analysis, we used an interior algorithm with augmented Lagrange multipliers \cite{ye1987thesis,rsolnp_package}. 

We conclude this section by noting that the theoretical parameter estimated by $\omega_n$ is \[
    \omega_{0n} := \mbox{argmax}_\omega \biggl\{ 1 - \frac{\frac{1}{K^\ast}\sum_{k=1}^{K^\ast} \mse_{0,\omega}(\Psi_\omega(T^{\circ}_{n,k}))}{\mse_{0,\omega}(\mu_{0,\omega})} \biggr\} \ . 
\]
We index this parameter by $n$ to denote that it is data-adaptive in the sense that it is unknown without first using the data to construct the super learners \cite{hubbard2016statistical}. 

\subsection{Estimating associations}
\begin{figure}
\centering
\includegraphics[width=0.7\textwidth]{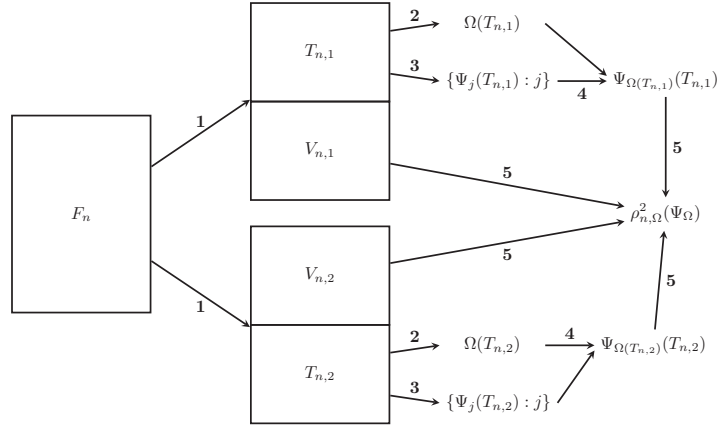}
\caption{A flowchart illustrating computation of two-fold cross-validated measure of association between $X$ and $Y$. This procedure maps the full data set into $\rho^2_{n,\Omega}(\Psi_\Omega)$, a real-number that measures the strength of association. Notation: $T_{n,k}$, the $k$-th training sample; $V_{n,k}$, the $k$-th validation sample; $\Psi_j(T_{n,k})$, the super learner for the $j$-th outcome fit in the $k$-th training sample; $\Omega(T_{n,k})$, outcome weights computed in the $k$-th training sample; $\Psi_{\Omega(T_{n,k})}(T_{n,k})$, the super learner fits from the $k$-th training sample combined using weights also computed in the $k$-th training sample; $\rho^2_{n,\Omega}(\Psi_{\Omega})$, the cross-validated nonparametric $R^2$ for the composite super learner.}
\end{figure}

We may report the maximized value $\rho^2_{n,\omega_n}(\Psi_{\omega_n})$ based on the full data $F_n := \{1,\dots,n\}$ as a relevant summary of the association between $X$ and $Y$. However, this estimate may be overly optimistic due to overfitting in the weight selection, particularly in situations when $Y$ is high-dimensional. Nevertheless, if one finds that $\rho^2_{n,\omega_n}(\Psi_{\omega_n})$ is less than or equal to zero, then it likely reasonable to conclude that there is no meaningful association between $X$ and $Y$. Otherwise, we propose to utilize an outer layer of $K$-fold cross-validation to accurately assess the association. This proposal is illustrated in Figure 3 for $K=2$. 

\noindent\textbf{Step 1. Partition the data. }
Starting with the full data set $\{O_i : i \in F_n\}$, randomly partition the data into $K$ splits of approximately equal size. As above, we use $T_{n,k} \subset F_n$ to denote the indices of observations in the $k$-th training sample and $V_{n,k}$ to denote the indices of observations in the $k$-th validation sample for $k=1,\dots,K$.

\noindent\textbf{Step 2. Fit super learner for each outcome in training data. }
For each split $k$, we use the training sample to fit each outcome, resulting in learner fits $\Psi_j(T_{n,k})$ for $j=1,\dots,J$. That is, we repeat the super learning procedure outlined in the previous subsection with $F_n^* = T_{n,k}$ for $k=1,\dots,K$. 

\noindent\textbf{Step 3. Fit outcome weights in training data. }
In the previous section, we wrote the procedure that finds outcome weights as a mapping $\Omega$ from a subset of observations to a vector of numbers. We now apply this procedure to the training data. That is, we repeat the outcome weighting procedure from the previous subsection with $F_n^\circ = T_{n,k}$ for $k=1,\dots,K$. 

\noindent\textbf{Step 4. Combine super learners based on outcome weights. }
Given weights $\Omega(T_{n,k})$, we compute the super learner prediction function for $Y_\Omega(T_{n,k})$ as $\Psi_{\Omega(T_{n,k})}(T_{n,k}) := \sum_{j=1}^J \Omega_j(T_{n,k}) \Psi_j(T_{n,k})$. 

\noindent\textbf{Step 5. Compute cross-validated nonparametric $R^2$. }
We use each validation sample to compute the mean squared-error for the combined super learner, $$
\mse_{n,\Omega,k}(\Psi_{\Omega}) := \frac{1}{|V_{n,k}|} \sum_{i \in V_{n,k}} \{ Y_{\Omega(T_{n,k}),i} - \Psi_{\Omega(T_{n,k})}(T_{n,k})(X_i) \}^2 \ ,
$$ 
which provides an evaluation of the performance of the composite super learner for predicting the composite outcome based on the weights learned in the training sample. We define $\mse_{n,\Omega}(\Psi_{\Omega}) := \frac{1}{K} \sum_{k=1}^{K} \mse_{n,\Omega,k}(\Psi_{\Omega})$ as the average of this measure across the $K$ splits. We evaluate the fit of the composite empirical mean $$
\bar{\Psi}_{\Omega(T_{n,k})}(T_{n,k}) := \frac{1}{|T_{n,k}|} \sum_{i \in T_{n,k}} Y_{\Omega(T_{n,k}) , i} \ , 
$$
for the composite outcome via $$
\mse_{n,\Omega,k}(\bar{\Psi}_{\Omega}) := \frac{1}{|V_{n,k}|} \sum_{i \in V_{n,k}} \{ Y_{\Omega(T_{n,k}),i} - \bar{\Psi}_{\Omega(T_{n,k})}(T_{n,k})\}^2 \ .
$$
Similarly as above, we define $\mse_{n,\Omega}(\bar{\Psi}_{\Omega}) := \sum_{k=1}^{K} \mse_{n,\Omega,k}(\bar{\Psi}_{\Omega})$ as the average across cross-validation folds. Finally, we compute the estimated cross-validated nonparametric $R^2$ measure of association as $$
\rho^2_{n,\Omega}(\Psi_{\Omega}) := 1 - \frac{\mse_{n,\Omega}(\Psi_{\Omega})}{\mse_{n,\Omega}(\bar{\Psi}_{\Omega})} \ .
$$

We conclude by noting that $\rho^2_{n,\Omega}(\Psi_{\Omega})$ estimates the data-adaptive parameter $$
\rho^2_{0n,\Omega}(\Psi_{\Omega}) := 1 - \frac{\sum_{k=1}^{K} \mse_{0,\Omega(T_{n,k})}(\Psi_{\Omega(T_{n,k})}(T_{n,k}))}{\sum_{k=1}^{K} \mse_{0,\Omega(T_{n,k})}(\bar{\Psi}_{\Omega(T_{n,k})}(T_{n,k}))} \ ,
$$
which describes the true cross-validated performance of the estimated composite prediction function for predicting the composite outcome. 

\subsection{Inference}

An asymptotically justified confidence interval can be constructed for $\rho^2_{n,\Omega}$ using influence function-based methodology \cite{hubbard2016statistical}. We propose to construct a Wald-style $100(1-\alpha)\%$ confidence interval about the estimate $\mbox{log}\{\mse_{n,\Omega}(\Psi_{\Omega})/\mse_{n,\Omega}(\bar{\Psi}_{\Omega})\}$ and back-transform this interval to obtain an interval on the original scale. Thus, our interval is of the form $$
1 - \mbox{exp}\biggl[\mbox{log}\biggl\{\frac{\mse_{n,\Omega}(\Psi_{\Omega})}{\mse_{n,\Omega}(\bar{\Psi}_{\Omega})}  \biggr\} \pm z_{1-\alpha/2} \frac{\sigma_n}{n^{1/2}} \biggr] \ ,
$$ 
where $z_{1-\alpha/2}$ is the $1-\alpha/2$ quantile of the standard normal distribution and $\sigma_n^2$ is a consistent estimate of the asymptotic variance of $\mbox{log}\{\mse_{n,\Omega}(\Psi_{\Omega})/\mse_{n,\Omega}(\bar{\Psi}_{\Omega})\}$. Similarly, a level $\alpha$ one-sided test of the null hypothesis of no association between $X$ and $Y$ can be performed by rejecting the null hypothesis whenever $$ \frac{\mbox{log}\{\mse_{n,\Omega}(\Psi_{\Omega})/\mse_{n,\Omega}(\bar{\Psi}_{\Omega})\}}{\sigma_n / n^{1/2}} < z_{\alpha} \ . $$

A closed-form variance estimator is constructed as follows. For $k=1,\dots,K^{\circ}$, we define \begin{align*}
D_{n,k}(\Psi_{\Omega})(O) &:= \{Y_{\Omega(T_{n,k})} - \Psi_{\Omega(T_{n,k})}(T_{n,k})(X) \}^2 - \mse_{n,\Omega,k}(\Psi_{\Omega}) \ , \ \mbox{and} \\
D_{n,k}(\bar{\Psi}_{\Omega})(O) &:= \{Y_{\Omega(T_{n,k})} - \bar{\Psi}_{\Omega(T_{n,k})}(T_{n,k})(X) \}^2 - \mse_{n,\Omega,k}(\bar{\Psi}_{\Omega}) \ . 
\end{align*}
These equations represent cross-validated estimates of the influence function of $\mse_{n,\Omega,k}(\Psi_{\Omega})$ and $\mse_{n,\Omega,k}(\bar{\Psi}_{\Omega})$, respectively, and we define $I_{n,k} := (D_{n,k}(\Psi_{\Omega}), D_{n,k}(\bar{\Psi}_{\Omega}))^T$ as the bivariate estimated influence function. We use the delta method to obtain a cross-validated estimate of the influence function of $\mbox{log}\{\mse_{n,\Omega}(\Psi_{\Omega})/\mse_{n,\Omega}(\bar{\Psi}_{\Omega})\}$. The estimated gradient of this transformation at the parameter estimates is $g_{n} := (\mse_{n,\Omega}^{-1}(\Psi_{\Omega}), -\mse_{n,\Omega}^{-1}(\bar{\Psi}_{\Omega}))^T$. A consistent variance estimator is  \begin{align*}
\sigma_n^2 := g_{n}^T \biggl\{ \frac{1}{K} \sum_{k=1}^{K} \frac{1}{|V_{n,k}|} \sum_{i \in V_{n,k}} I_{n,k}(O_i)I_{n,k}^T(O_i) \biggr\} g_{n} \ . 
\end{align*}

\section{Variable importance}

Often we are not only be interested in an overall summary of the association between $X$ and $Y$, but also in the relative importance of each component of $X$. Ensemble machine learning may be criticized as a ``black box'' approach, in which it is difficult to understand the relative contributions of different variables.\cite{freitas2014comprehensible} To provide better understanding of the ``black box,'' we propose to study differences in the estimated association between $X$ and $Y$ when considering different subsets of variables. Our proposal is similar to variable importance measures proposed for specific machine learning algorithms, such as random forests,\cite{breiman2001random} because they measure the change in predictive performance with and without each predictor variable considered. Although existing approaches may have poorly behaved statistical inference,\cite{strobl2008danger} the present approach yields straightforward, asymptotically justified inference.

To assess the importance of a subset $S \subset \{1,\dots, D\}$ of variables $\tilde{X} := \{X_{s} : s \in S\}$, we propose to repeat our procedure, but restrict only to variables not in this subset. We obtain an estimate $\rho^2_{n,\tilde{\Omega}}(\tilde{\Psi}_{\tilde{\Omega}})$ of the cross-validated performance measure for the composite prediction algorithm based on this subset of variables. The joint importance of the variables in $\tilde{X}$ is defined by a contrast between $\rho^2_{0n,\Omega}(\Psi_{\Omega})$ and $\rho^2_{0n,\tilde{\Omega}}(\Psi_{\tilde{\Omega}})$ such as \begin{equation} 
\Delta_{0n}(\Psi_{\Omega}, \tilde{\Psi}_{\tilde{\Omega}}) := \rho^2_{0n,\Omega}(\Psi_{\Omega}) - \rho^2_{0n,\tilde{\Omega}}(\Psi_{\tilde{\Omega}}) \ . \label{varImportanceParameter}
\end{equation}
Point estimates for (\ref{varImportanceParameter}) are constructed by plugging in the estimates, and asymptotically justified confidence intervals and hypothesis tests about these estimates are constructed using influence functions as in the previous section. 
    
%The optimal weighting scheme for the outcomes may change when different sets of variables are considered. Studying whether and how the weights change with different sets of variables may provide important descriptive information for how the variables in $X$ are associated with the various outcomes. That is, exclusion of specific components of $X$ may not change the predictive performance for the composite outcome, but may change how the outcomes are combined. More directly, we may study the influence of components of $X$ on a particular $Y_j$ by fixing $\omega_j = 1$, setting all other weights to 0. Comparing changes in our measure of association for this choice of weights directly assesses how much these components of $X$ contribute to the overall association with $Y_j$. In fact, these estimates are computed as intermediate steps in our procedure and may be used to help determine the sources of the overall association between $X$ and $Y$.  

\section{Simulations}

We evaluated the finite-sample performance of the proposed estimators in two simulations. In the first, we studied the operating characteristics of our estimator. This simulation confirms that the asymptotic theory developed in previous sections leads to reasonable finite-sample performance for the estimators. In the second simulation, we compared our proposed method to existing methods for assessing associations between $X$ and $Y$. This simulation shows that our method correctly identifies associations in situations where existing methods do not. 

\subsection{Simulation 1: operating characteristics}
\begin{figure}
\centering
\includegraphics[scale=0.75]{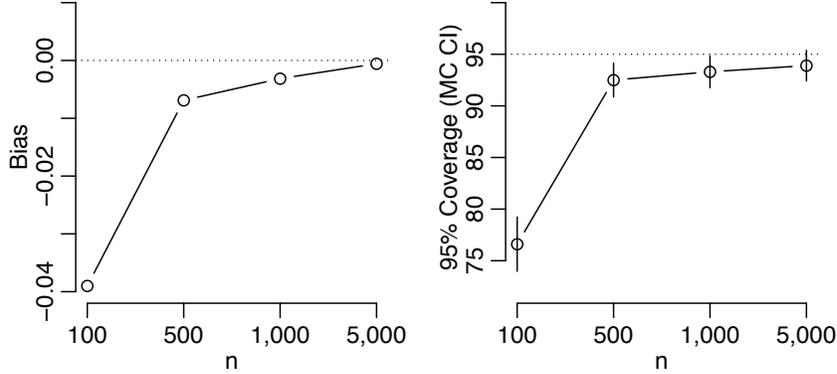}
\caption{Results of the first simulation study. Left: Bias of $\rho^2_{n,\Omega}(\Psi_\Omega)$. Right: Estimated coverage of nominal 95\% intervals (with Monte Carlo confidence intervals) for the proposed confidence interval.}
\label{sim1Rslt}
\end{figure}

We generated 1000 simulated data sets by independently sampling $n$ copies of $X_1,X_2, X_3, X_7, X_8, X_9$ from a Uniform(0,4) distribution and $n$ copies of $X_4, X_5, X_6$ from a Bernoulli distribution with success probabilities of 0.75, 0.25, and 0.5. We sampled $\epsilon_1, \epsilon_2, \epsilon_3$ independently from a Normal(0,5) distribution and let \begin{align*}
Y_1 &= X_1 + 2X_2 + 4X_3 + X_4 + 2X_5 + 4X_6 + 2X_7 + \epsilon_1 \ , \\
Y_2 &= X_1 + 2X_2 + 4X_3 + X_4 + 2X_5 + 4X_6 + 2X_8 + \epsilon_2 \ , \ \mbox{and} \\
Y_3 &= X_1 + 2X_2 + 4X_3 + X_4 + 2X_5 + 4X_6 + 2X_9 + \epsilon_3 \ . 
\end{align*}
The optimal $R$-squared for each outcome is about 0.60. The true optimal weighting scheme for the outcomes is $\omega_0 = (1/3, 1/3, 1/3)$ and the true optimal $R$-squared for the optimally composite outcome is approximately 0.81. We considered sample sizes of $n=100,500,1000,5000$. For computational simplicity, we considered a small super learner library consisting of a main terms linear regression (function \texttt{SL.glm} in \texttt{SuperLearner} package), an intercept-only regression (\texttt{SL.mean}), and a forward-stepwise selection linear regression model (\texttt{SL.step.forward}). We used ten folds for each layer of cross-validation, $K = K^* = K^{\circ} = 10$. 

The estimates of predictive performance were biased downwards in smaller sample sizes (Figure \ref{sim1Rslt}). However, even with $n=100$ the average estimated performance was only about 5\% too small. The confidence interval coverage was less than nominal for small sample sizes, but had nominal coverage in larger samples. 

\begin{figure}
\centering
\includegraphics[scale=0.75]{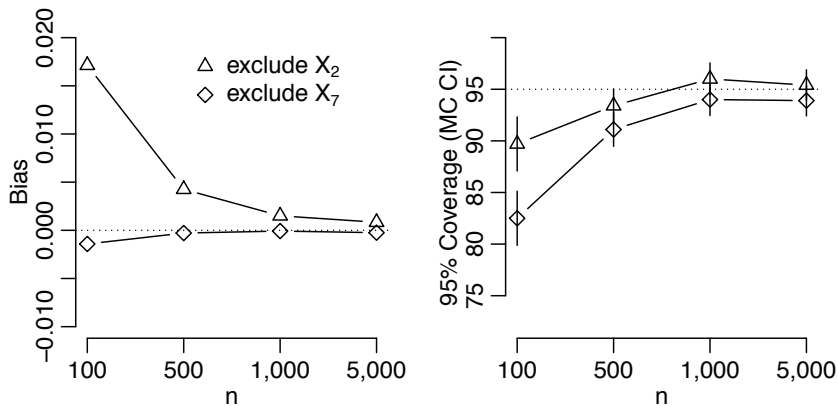}
\caption{Results of second simulation study.  Left: Bias of the estimated variable importance measure comparing a super learning that uses all of the variable to one that omits either $X_2$ (triangle) or $X_7$ (diamond). Right: Estimated coverage for nominal 95\% intervals (with Monte Carlo confidence intervals) for the proposed confidence intervals.}
\label{sim2Rslt}
\end{figure}

We excluded $X_2$ and then excluded $X_7$ to estimate the additive difference in $R$-squared values (equation (\ref{varImportanceParameter})) for these 2 variables. Examining the data-generating mechanism, we note that $X_2$ was important for predicting each individual outcome and should be important for predicting the composite outcome. The optimal $R$-squared for predicting each individual outcome without $X_2$ was 0.52, and the additive importance of $X_2$ was 0.60 - 0.52 = 0.08. For the composite outcome, the optimal weightings were unchanged $(1/3,1/3,1/3)$, but the optimal $R$-squared without $X_2$ decreased to 0.68 and the additive importance of $X_2$ for the composite outcome was 0.81 - 0.68 = 0.13. In contrast, $X_7$ was important only for predicting $Y_1$ and had no effect on predicting $Y_2$ or $Y_3$. Therefore, the optimal composite outcome was expected to upweight $Y_2$ and $Y_3$ when excluding $X_7$, and the importance of $X_7$ for predicting the optimal composite outcome may be minimal. The optimal weights without $X_7$ were $(0.24, 0.38, 0.38)$, whereas the optimal $R$-squared without $X_7$ was 0.79, leading to an additive importance of 0.81 - 0.79 = 0.02. 

The importance measures for $X_2$ exhibited substantial bias (25\% truth) when $n=100$, but both measures were unbiased with sample sizes $> 1,000$ (Figure \ref{sim2Rslt}). The nominal 95\% confidence interval coverage was less than nominal in small samples, but the coverage was $>90\%$ for sample sizes $ >500$.

\section{Simulation 2: power of associative hypothesis tests}

In this simulation, $X$ was simulated as above. The outcome $Y$ was simulated as a 10-dimensional multivariate normal variate with mean $\mu = (\mu_1(x), \dots, \mu_{10}(x))^T$ and covariance matrix $\Sigma$. Letting $\Sigma_{i,j}$ denote the $(i,j)$ entry in $\Sigma$, we set $\Sigma_{i,i} = 5$ for $i=1,\dots,10$. We also set $\Sigma_{1,2} = \Sigma_{1,3} = \Sigma_{4,7} = -2$ and $\Sigma_{3,9} = \Sigma_{4,9} = \Sigma_{5,9} = 2$, while also appropriately setting the corresponding lower triangle elements to these values. All other off-diagonal elements were set to zero. We studied three settings. The first setting had no association between $X$ and $Y$, and we set $\mu_j(x) = 0$ for $j=1,\dots,10$. The second setting had a linear association between $X$ and one component of $Y$. Here, we set $\mu_6(x) = -2 + 0.75 x_1$ and $\mu_j(x) = 0$ for $j \ne 6$. The third setting had a nonlinear association between $X$ and one component of $Y$. Here, we set $\mu_6(x) = -2 + 0.75(x-2)^2$ and $\mu_j(x) = 0$ for $j \ne 6$. 

We studied the power of level 0.05 tests of the hypothesis of no association between $X$ and $Y$. Power was estimated as 100 times the proportion of the 1,000 total simulations where the null hypothesis was rejected. We tested this hypothesis with the proposed one-sided hypothesis test with all layers of cross validation set to 5-fold. We used a super learner library that included an intercept-only regression, a main terms regression, and a generalized additive model.\cite{hastie1990generalized} We also tested this hypothesis using four common tests of canonical correlation: Wilks' $\Lambda$, the Hotelling-Lawley trace, the Pillai-Bartlett trace, and Roy's largest root. Rather than comparing these statistics to their respective approximate asymptotic distributions, we used permutation tests to compute p-values.\cite{ccpPackage} We also studied the power of a test based on a principal components approach. For this test, we constructed a composite outcome based on the first principal component of $Y$ and subsequently fit a main-terms linear regression on $X$. The null hypothesis of no association was rejected whenever the p-value associated with the global F-test for this linear regression was less than 0.05. 

\begin{figure}
\includegraphics[scale = 0.6]{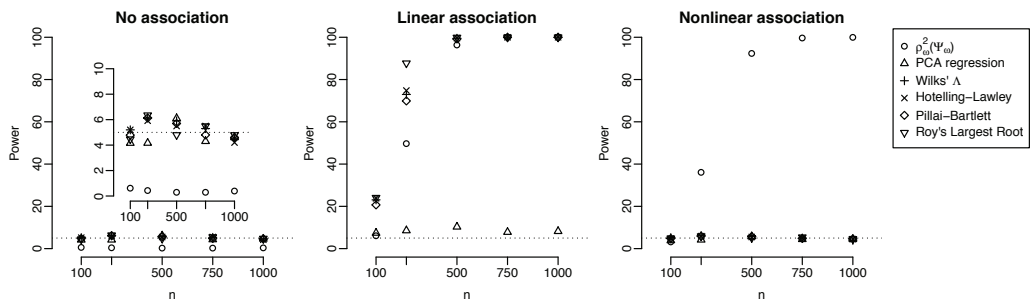}
\caption{Power of various level 0.05 tests for rejecting the null hypothesis of no association between $X$ and $Y$ under the null hypothesis (left), linear association (middle), and nonlinear association (right).}
\label{powerSimResultFig}
\end{figure}

Under the null hypothesis, each canonical correlation-based and the PCA-based test had approximately nominal type I error (Figure \ref{powerSimResultFig}, left panel). However, our proposed test was found to be conservative, falsely rejecting the null hypothesis less than 1\% of the time. In the second scenario, the canonical correlation-based tests had greater power to reject the null hypothesis than our proposed test at the two smallest sample sizes (middle panel). The PCA-based test had poor power, as the first principal component fails to adequately reflect the component of $Y$ for which a relationship with $X$ exists. In the third scenario, only our proposed test had power to correctly reject the null hypothesis (right panel).

\section{Data Analysis}
\begin{table}
\begin{tabular}{l|l}
\hline
\textbf{Group} & \textbf{Variables} \\ 
\hline
Health care & Health care access, use of preventive health care \\
Household & Child:adult ratio, child dependency ratio, crowding index, urban score \\
Socioeconomic status & Total income, socioeconomic status \\
Water and sanitation & Sanitation, access to clean water \\
Parental & Mother age, father age, mother height, mother education (y), \\
         & father education (y), marital status, mother age first child, parity \\
Growth & Weight-for-age Z-score, height-for-age z-score \cite{who2006growth} (0, 6, 12, 18, 24 mo) \\
Other & Mother smoked during pregnancy, child's sex, gestational age at birth \\
\hline
\end{tabular}
\caption{Variables used to predict achievement test scores in Cebu Longitudinal Health and Nutrition Study analysis.}
\label{cebuVariables}
\end{table}

The ongoing Cebu Longitudinal Health and Nutrition Study (CLHNS) enrolled Filipino women who gave birth in 1983-1984, and the children born to these women have been followed prospectively.\cite{clhnswebsite,adair2010cohort} The long-term follow-up with these children has enabled researchers to quantify the long-term effects of prenatal and early childhood nutrition and health on outcomes (adolescent and adult health, economics, and development).\cite{daniels2004growth,daniels2005breast} We are interested in assessing the strength of correlation between prenatal and early childhood data and later schooling achievement. We are also interested in understanding the extent to which somatic growth (height and body weight) early in life associates with later neurocognitive outcomes.\cite{carba2009early,smithers2013impact} In 1994, achievement tests were administered to 2166 children in 3 subjects: mathematics, and English and Cebuano languages. 

We applied the present methods to assess association of the 3 test scores with variables collected from birth to age two years (Table \ref{cebuVariables}). For variables that had missing values, we created an indicator of missingness, and missing values of the original variable were set = 0. The library of candidate super learner algorithms included a random forest, gradient boosted machines, and elastic net regression algorithms. The tuning parameters for each algorithm were chosen via nested 5-fold cross-validation. We used 10-fold cross-validation for each layer of cross-validation.

The estimated cross-validated R-squared for predicting test scores was 0.24 (95\% CI: 0.21, 0.27) for mathematics, 0.31 (95\% CI: 0.28, 0.34) for English, and 0.23 (95\% CI: 0.20, 0.26) for Cebuano. The estimated association with the composite outcome was 0.32 (95\% CI: 0.28, 0.35). The estimated association with the optimally weighted outcome was only slightly higher than predicting an equally weighted outcome (0.30, 95\% CI: 0.27, 0.34), as well as an outcome weight based on a linear combination based on the first principal component of the test scores (0.28, 95\% CI: 0.25, 0.31).   

\begin{figure}
\centering
\includegraphics[scale=0.75]{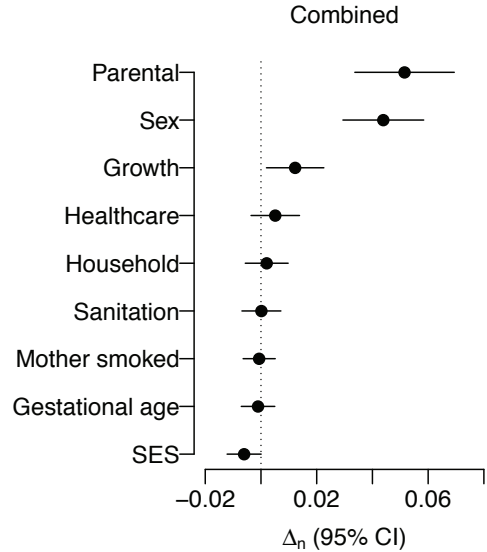}
\caption{The change in $R$-squared for predicting each outcome and the composite outcome when eliminating groups of variables. Abbreviations, CI, confidence interval; SES, socioeconomic status.}
\label{groupVariables}
\end{figure}

We computed variable importance measures by repeating the procedure, eliminating groups of variables and estimating the additive change in performance. We found that the child's sex and parental information were responsible for the largest proportion of the association with achievement test score performance (Figure \ref{groupVariables}). However, the somatic growth variables, as a group, modestly increased the association, with an estimated change in association of 0.01 (95\% CI: 0.00, 0.02; $P$ = .02). 

\section{Discussion}

Our proposed method provides a new means of summarizing associations with multivariate outcomes and our simulations demonstrate that it provides powerful inference for tests of association when complex and nonlinear relationships are present. We found that existing tests provide greater power in small samples against alternatives with linear relationships between covariates and outcomes. This is unsurprising as our approach relies on several layers of cross-validation, which may stretch small samples too thin. We also found that our test was conservative under the null hypothesis. This can be explained by the fact that the true value of cross-validated $R$-squared was often less than zero, while our test was based on testing a value of zero for this parameter. We suggest that a permutation test could be used to construct a hypothesis test with better operating characteristics; however, this may often prove to be computationally intractable in practice. In such cases, we may instead opt for the more conservative, but less computationally intensive test. 

A possible criticism of our approach is the data-adaptive nature of the parameter $\rho^2_{0n,\Omega}(\Psi_{\Omega})$. One could argue that this parameter is only a useful measure of association insofar as the relevant prediction functions approximate $\psi_{0,\omega_0}$, the unknown true conditional mean of the optimally combined outcome. This strengthens the argument for using super learning to a prediction function that is as accurate as possible. A more direct measure of association is (\ref{trueR2}) evaluated at $\omega_0$. Future work will be devoted to strategies for direct estimation of this quantity. On the other hand, as mentioned above, in some settings the goal is to construct a composite outcome and a prediction function for that outcome. In these cases, our associative measure is directly relevant, as it provides an approximation to how well we might expect to predict the composite outcome of future patients. 

Though our proposal is computationally intensive, the vast majority of the computational burden lies in fitting the candidate learners, which can be implemented in a parallelized framework. The computational burden of the procedure may be further reduced by clever choices of number of cross-validation folds. When $K = K^* = K^\circ$, the procedure requires fitting $K^3 \sum_{j=1}^J M_j$ total candidate learners. However, if $K^\circ = K-1$ and $K^* = K-2$, it is possible to re-use candidate learner fits at different points in the procedure, thereby drastically decreasing the computational burden of the procedure. In fact, we can show that this approach requires only $(K^3 + 5K) / 6 \times \sum_{j=1}^J M_j$ candidate learner fits. Thus, for $K = 10$, the latter approach is expected to execute in 17.5\% the time it would take if $K = K^* = K^\circ = 10$. An R package implementing our methods is available (link to be added upon acceptance).  

% Future extensions may consider nonlinear combinations of the outcome, perhaps via alternating conditional expectation.\cite{breiman1985estimating} It may also be of interest to consider extensions for estimating covariate-adjusted effects of variables on data-adaptively-combined outcome; estimation of these effects and associated inference may be facilitated via cross-validated targeted maximum likelihood.\cite{zheng2010asymptotic,hubbard2016statistical}

\bibliographystyle{elsarticle-num} 
\bibliography{myreferences}
\end{document}